\pdfoutput=1
\documentclass[10pt,conference]{IEEEtran}

\usepackage{graphicx}
\usepackage{tabularx}

\usepackage[T1]{fontenc}
\usepackage[utf8]{inputenc}

\usepackage{csquotes}
\usepackage{nicefrac}
\usepackage[textsize=tiny]{todonotes}
\usepackage{booktabs}
\usepackage{xcolor}
\usetikzlibrary{arrows,positioning} 
\usetikzlibrary{shapes.geometric, positioning, quantikz}
\usepackage{nicematrix}
\usepackage{flushend}
\usepackage{hyperref}
\usepackage[binary-units=true, detect-all=true]{siunitx}
\usepackage{etoolbox}
\usepackage{soul}
\usepackage{amssymb}

\robustify\bfseries

\makeatletter
\let\MYcaption\@makecaption
\makeatother
\usepackage[font=footnotesize]{subcaption}
\makeatletter
\let\@makecaption\MYcaption
\makeatother

\usepackage{yquant}
\usepackage{pgfplots}

\usepackage[style=ieee, maxnames=3, minnames=1, date=year, doi=false,isbn=false,backend=biber]{biblatex}
\usepackage[binary-units=true, detect-all=true]{siunitx}
\usepackage{etoolbox}
\robustify\bfseries

\usepackage{nicematrix}
\usepackage{flushend}
\usetikzlibrary{fit, quotes}
\usepackage{bbm}

\usepackage{amsthm}

\newtheorem{example}{Example}

\AtEveryBibitem{
  \clearname{editor}%
  \clearfield{series}%
  \clearfield{isbn}%
  \clearfield{issn}%
  \clearfield{volume}%
  \clearfield{number}%
  \clearfield{pages}%
}

\begin{document}

\title{Towards an Automated Framework for\\ Realizing Quantum Computing Solutions}

\author{
	\IEEEauthorblockN{Nils Quetschlich\IEEEauthorrefmark{1}\hspace*{1.5cm}Lukas Burgholzer\IEEEauthorrefmark{2}\hspace*{1.5cm}Robert Wille\IEEEauthorrefmark{1}\IEEEauthorrefmark{3}}
	\IEEEauthorblockA{\IEEEauthorrefmark{1}Chair for Design Automation, Technical University of Munich, Germany}
	\IEEEauthorblockA{\IEEEauthorrefmark{2}Institute for Integrated Circuits, Johannes Kepler University Linz, Austria}
	\IEEEauthorblockA{\IEEEauthorrefmark{3}Software Competence Center Hagenberg GmbH (SCCH), Austria}
	\IEEEauthorblockA{\href{mailto:nils.quetschlich@tum.de}{nils.quetschlich@tum.de}\hspace{1.5cm}\href{mailto:lukas.burgholzer@jku.at}{lukas.burgholzer@jku.at}\hspace{1.5cm} \href{mailto:robert.wille@tum.de}{robert.wille@tum.de}\\
	\url{https://www.cda.cit.tum.de/research/quantum}
	}
}

\maketitle

\begin{abstract}
Quantum computing is fast evolving as a technology due to recent advances in hardware, software, as well as the development of promising applications.
To use this technology for solving specific problems, a suitable quantum algorithm has to be determined, the problem has to be encoded in a form suitable for the chosen algorithm, it has to be executed, and the result has to be decoded.
To date, each of these tedious and error-prone steps is conducted in a mostly manual fashion.
This creates a high entry barrier for using quantum computing---especially for users with little to no expertise in that domain.
In this work, we envision a framework that aims to lower this entry barrier by allowing users to employ quantum computing solutions in an automatic fashion.
To this end, interfaces as similar as possible to classical solvers are provided, while the quantum steps of the workflow are shielded from the user as much as possible by a fully automated backend.
To demonstrate the feasibility and usability of such a framework, we provide \mbox{proof-of-concept} implementations for two different classes of problems which are publicly available on GitHub (\url{https://github.com/cda-tum/MQTProblemSolver}) as part of the Munich Quantum Toolkit (MQT).
By this, this work provides the foundation for a \mbox{low-threshold} approach realizing quantum computing solutions with no or only moderate expertise in this technology.
\end{abstract}

\section{Introduction}\label{sec:introduction}
Quantum computing devices are rapidly evolving and maturing with the increase of the number of available quantum computers as well as their number of qubits,
error rates decreasing, and operations becoming faster. 
In parallel, numerous \emph{Software Development Kits}~(SDKs), such as Google's Cirq~\cite{cirq}, IBM's~Qiskit \cite{qiskit}, Quantinmuum's TKET~\cite{sivarajahKetRetargetableCompiler2020}, and Rigetti's Forest~\cite{rigetti}, are being developed to make use of the available quantum computing hardware.
Even specialized SDKs for certain purposes are available, e.g., Xanadu's Pennylane~\cite{pennylane} for differentiable quantum computing.
These developments spark interest in quantum computing from academia and industry---leading to potential applications in various domains such as physics~\cite{wf_ex_1}, chemistry~\cite{wf_ex_2}, finance~\cite{wf_ex_3}, and optimization~\cite{wf_ex_4}.

So far, many works aiming to solve specific problems by utilizing quantum computing follow a similar workflow consisting of four steps:
\begin{enumerate}
\item Selecting a suitable quantum algorithm.
\item Encoding the specific problem into a quantum circuit.
\item Executing it on a quantum device.
\item Decoding the solution from the quantum result.
\end{enumerate}

While this has led to several promising quantum computing applications (triggering a substantial momentum for quantum computing in general), realizing the respective solutions comes with two major challenges:
First, for all four steps, expertise in quantum computing is required. 
Without that, neither a quantum algorithm can be selected if the user is not aware of its prerequisites, nor can the problem be encoded, or the resulting quantum circuit be executed and the solution be extracted.
Naturally, most of the users from those application domains are not trained experts in quantum computing which poses a huge roadblock in the further utilization and adoption of quantum computing.
Second, especially during the encoding and decoding, many tedious and \mbox{error-prone} tasks have to be conducted---resulting in a huge manual effort to actually solve problems using quantum computing.
Both aspects combined lead to a high entry barrier to employ quantum computing and make its utilization very challenging.

In this work, we envision a framework that simplifies the realization of quantum computing solutions---particularly for users from the various application domains. 
To this end, we exploit the fact that the current workflow summarized above actually offers tangible opportunities to shield the user as much as possible from the intricacies of quantum computing. 
This is accomplished by keeping the interfaces for both, the problem input and the solution output formats, as similar as possible to classical solvers
and by providing guidance for the quantum algorithm selection procedure. 
Using this as a basis, the remaining steps (encoding, executing, and decoding) are then covered in a fully automated fashion.
 
To demonstrate the feasibility and usability of such a framework, a \mbox{proof-of-concept} implementation---which is publicly available on GitHub (\url{https://github.com/cda-tum/MQTProblemSolver}) as part of the \emph{Munich Quantum Toolkit (MQT)}---has been realized for two different problem classes: \emph{Satisfiability Problems} (SAT problems) and \emph{Graph-based Optimization Problems}. For both, corresponding case studies confirmed the benefits from a user's perspective. By this, this work provides the foundation for a \mbox{low-threshold} approach of realizing quantum computing solutions with no or only moderate background in this technology.

\vspace{10cm}
\noindent The rest of this work is structured as follows: \autoref{sec:background} gives a short introduction to quantum computing. 
Based on that, a detailed explanation of the mentioned workflow of realizing quantum computing solutions for specific problems is given in \autoref{sec:workflow}.
Afterwards, the tangible opportunities to automate and simplify this workflow are identified in \autoref{sec:framework}---motivating the envisioned framework.
Based on that, the \mbox{proof-of-concept} implementations are described in \autoref{sec:implementation} and evaluated from a user's perspective in \autoref{sec:results}. \autoref{sec:conclusions} concludes this work.

\vspace{-1mm}
\section{Quantum Computing}\label{sec:background}
In order to keep this work self-contained, this section gives a short introduction to quantum computing.
Compared to classical computing, where each bit can have a value of $0$ or $1$ representing its state, a quantum bit or \emph{qubit} may also be in a \emph{superposition} of those values, i.e., the \emph{quantum state} $\ket{\phi}$ of a qubit can be written as
\[
\ket{\phi}= \alpha_0 \ket{0} + \alpha_1 \ket{1} =   \begin{bmatrix}
    \alpha_0 \:	
    \alpha_1 
  \end{bmatrix}^T
\]
with amplitudes $\alpha_0$, $\alpha_1 \in \mathbb{C}$ such that $ |\alpha_0|^2 + |\alpha_1|^2 =1$.
For $n$ qubits, their state is composed of $2^n$ amplitudes $\alpha_i \in \mathbb{C}$ with $i$ from $0$ to $2^n -1$.
Again, the quantum state can be written as a superposition of all its basis states, i.e.,
\[
\ket{\phi}=  \sum_{i=0}^{2^n -1} \alpha_i \ket{i} =
\begin{bmatrix}
    \alpha_0 \:
    \hdots \:
    \alpha_{2^n -1}
  \end{bmatrix}^T \text{with} \sum_{i} |\alpha_i|^2=1.
  \]

Analogously to classical computing and its logical gate operations, computations on quantum computers are conducted using quantum gates which alter the state of the qubit. The corresponding functionality of a quantum gate can be described by a \emph{unitary} matrix; its effect on a quantum state can be determined by multiplying the matrix representation and the currently considered state representation. 

Three prominent gates acting on a single qubit are the \emph{Hadamard} (H), the \emph{Pauli-X} (X), and the \emph{Pauli-Z} (Z) gate which are defined by the matrices
\[
H= \frac{1}{\sqrt{2}}
  \begin{bmatrix}
    1 & 1 \\
    1 & -1 
  \end{bmatrix}, \: 
  X=
  \begin{bmatrix}
    0 & 1 \\
    1 & 0 
  \end{bmatrix},\:
    \text{and }
  Z=
  \begin{bmatrix}
    1 & 0 \\
    0 & -1 
  \end{bmatrix}.
  \]
  A prominent representative acting on two qubits (typically referred to as \emph{control} and \emph{target} qubits) is the \mbox{controlled-not}~(CNOT) gate.
  If the control qubit is $\ket{1}$, the CNOT gate switches the amplitudes of the target qubit.
In principle, any operation can be controlled by arbitrarily many qubits. 
A corresponding multi-controlled operation is only applied, if all control qubits are $\ket{1}$.
  An example for such a multi-controlled operation is the \emph{\mbox{multi-controlled}-Z} (MCZ) gate.
  If all the control qubits are~$\ket{1}$, the MCZ gate applies a Z gate to the target qubit.
  Those two operations with the second respectively the $n^\mathit{th}$~qubit being the target qubit are defined by the matrices
  \[
  \mathit{CNOT} = 
    \begin{bmatrix}
    1 & 0 & 0 & 0  \\
    0 & 1 & 0 & 0  \\
    0 & 0 & 0 & 1  \\
    0 & 0 & 1 & 0 
  \end{bmatrix}
    \text{ and }
    \mathit{MCZ} = 
    \begin{bmatrix}
    1 & 0  &  \hdots & 0 \\
    0 &  \ddots &  \ddots & \vdots  \\
    \vdots &  \ddots & 1   &0\\
    0 &  \hdots & 0  & -1
  \end{bmatrix}.
\]

The state of a quantum system cannot be directly observed.
Instead, a \emph{measurement} collapses the state to one of the (classical) basis states $\ket{i}$---each with probability $|\alpha_i|^2$---which can then be read out.

Quantum algorithms are typically described in the form of a \emph{quantum circuit}, i.e., a sequence of gates that are applied to the qubits of a quantum system.

\begin{figure}[t]
\centering
   \resizebox{0.4\linewidth}{!}{
				\begin{tikzpicture}
				  \begin{yquant}	
						qubit {$\ket{0} $} q[+1];
						qubit {$\ket{0} $} q[+1];
										    
				    	box {$H$} q[0];
				    	cnot q[1] | q[0];
				    	
				    	measure q[0];
				    	measure q[1];
				    	
				  \end{yquant}
				\end{tikzpicture}}
		\caption{Exemplary quantum circuit starting in state $\ket{00}$.}
         \label{fig:circuit}
\vspace{-5mm}
\end{figure}
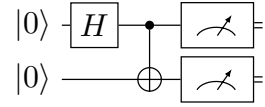

\begin{example}
\autoref{fig:circuit} shows an exemplary quantum circuit with two qubits that first applies a Hadamard gate, followed by a CNOT gate.
In the end, the qubits are measured.
\end{example}

\section{Realizing Quantum Computing Solutions}\label{sec:workflow}

This section gives an overview of the main steps to be conducted when aiming to solve a specific problem using quantum computing.
Based on that, the remainder of this work will then deal with how to automate this workflow or, at least, aid the user during this process.
In order to properly illustrate those steps as well as the proposed solution, a running example is used, which is introduced in the following.

\begin{figure*}[h]
\begin{subfigure}{.12\linewidth}
    \centering
        	\begin{tabular}{c|cc}
$\sum$  & $1$ & $1$    \\ \hline
$1$  & $a$  & $b$   \\ 
$1$  & $c$  & $d$   \\ 
\end{tabular}
\end{subfigure}
{$\xrightarrow{Enc.}$}
\begin{subfigure}{.43\linewidth}
    \centering
   \resizebox{1.0\linewidth}{!}{
				\begin{tikzpicture}
				  \begin{yquant*}

						qubit {$a$} q;
						qubit {$b$} q[+1];
						qubit {$c$} q[+1];
						qubit {$d$} q[+1];
						qubit {$flag$} q[+1];
						
						box {Prep.} (q);
						
						[this subcircuit box style={dashed, rounded corners, inner ysep=4pt,  "$\times2$"}]
						subcircuit {
						qubit {} q[5];
						[this subcircuit box style={rounded corners, inner ysep=4pt,  "Oracle"}]
						subcircuit {
						qubit {} q[5];
				    	cnot q[2] | q[3];
				    	text {$c \neq d$} q[2];
				    	cnot q[3] | q[1];
				    	text {$b \neq d$} q[3];
				    	cnot q[1] | q[0];
				    	text {$a \neq b$} q[1];
				    	zz (q[1-4]);
				    	cnot q[1] | q[0];
				    	cnot q[3] | q[1];
				    	cnot q[2] | q[3];
						} (q[0-4]);			
						
						box {Diff.} (q[0-3]);	
						} (q[0-4]);			
						
				    	measure (q[0-3]);
 				    	
				  \end{yquant*}
				\end{tikzpicture}}
\end{subfigure}
{$\xrightarrow{Exec.}$}
\begin{subfigure}{.18\linewidth}
    \centering
   \resizebox{1.0\linewidth}{!}{
\begin{tikzpicture}
    \begin{axis}
        [
        ybar,
        ymax=0.5,ymin=0,
        font=\LARGE,
        xticklabel style={rotate=90},
        ,ylabel=Relative Frequency
       ,xtick={0,1,2,3,4,5,6,7,8,9,10}
        ,xticklabels={$\ket{0000}$, , $\ket{0101}$,$\ket{0110}$,$\ket{0111}$,,$\ket{1000}$,$\ket{1001}$,$\ket{1010}$,,$\ket{1111}$}
        ]
        \addplot coordinates
        {(0,0.004)(1,0.0)  (2,0.004) (3,0.47) (4,0.004)(5,0.0)  (6,0.004) (7,0.47)(8,0.004)(9,0.0)  (10,0.004) };
    \end{axis}
    
	\node[align=center] at (1.1,-1.0) {\huge{$...$}};		
	\node[align=center] at (3.4,-1.0) {\huge{$...$}};		
	\node[align=center] at (5.7,-1.0) {\huge{$...$}};		
	
\end{tikzpicture}

}
\end{subfigure}
{$\xrightarrow{Dec.}$}
\begin{subfigure}{.10\linewidth}
    \centering
            	\begin{tabular}{c|cc}
$\sum$   & $1$ & $1$    \\ \hline
$1$  & $0$  & $1$   \\ 
$1$  & $1$  & $0$   \\ 
\end{tabular}

\vspace{3mm}

\begin{tabular}{c|cc}
$\sum$   & $1$ & $1$    \\ \hline
$1$  & $1$  & $0$   \\ 
$1$  & $0$  & $1$   \\ 
\end{tabular}
\end{subfigure}

	\subfloat[Initial problem.\label{fig:wf1}]{\hspace{.15\linewidth}}
	\hspace{2.6cm}
	\subfloat[Quantum circuit. \label{fig:wf2} ]{\hspace{.2\linewidth}}
	\hspace{2.7cm}
	\subfloat[Histogram. \label{fig:wf3} ]{\hspace{.2\linewidth}}
	\hspace{2mm}
	\subfloat[Solution.\label{fig:wf4} ]{\hspace{.12\linewidth}}

\caption{Workflow from an initial problem instance to a valid solution using quantum computing.}
\label{fig:workflow}
\end{figure*}
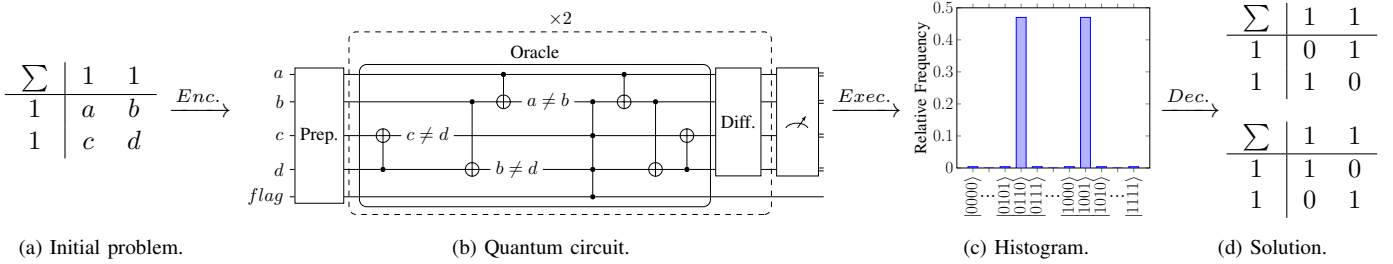

\begin{example}\label{ex:kakuro}
Kakuro, a cross-sum riddle, is an example of a SAT problem and is composed of a grid structure with $M$ rows and $N$ columns, where each row and column shall add up to a given sum.
Additionally, numbers within each row and each column must be distinct. 
A simple Kakuro riddle instantiation with a grid structure of $2\times2$ is shown in \autoref{fig:wf1}, where the goal is to determine $a$, $b$, $c$, and $d$ such that the respective sums add up to~$1$.
Thus, all those variables are to be assigned %
either $0$ or $1$.
A solution to this problem is characterized by satisfying the constraints $a \neq b$, $b \neq d$, and $c \neq d$.
\end{example}

\subsection{Quantum Algorithm Selection}\label{sec:algos}
The first step towards solving a problem using quantum computing is to choose a proper quantum algorithm which is suitable for the considered problem.
Without going into the details of quantum algorithms (for this, we refer to the given references), some prominent representatives are

\begin{itemize}
\item \emph{Grover's Algorithm}~\cite{grover}, which is suited for unstructured search (problems),
\item \emph{Quantum Phase Estimation} (QPE, \cite{kitaevQuantumMeasurementsAbelian1995}), which provides the foundation of \emph{Shor's Algorithm}~\cite{shor} used for integer factorization, or
\item \emph{Variational Quantum Algorithms} (VQAs, \cite{cerezoVariationalQuantumAlgorithms2020}), e.g., \emph{Variational Quantum Eigensolver} (VQE) and \emph{Quantum Approximate Optimization Algorithm} (QAOA), which allows for hybrid classical-quantum computing while also being suited for combinatorial optimization problems, such as the \emph{\mbox{Max-Cut} Problem}.
\end{itemize}

\begin{example}\label{ex:algo_choice}
Finding a solution for a SAT problem such as the Kakuro riddle can be interpreted as finding an assignment of all variables subject to some constraints.
Being an unstructured search problem, Grover's algorithm is a very suitable choice for this. 
The corresponding quantum circuit is composed of three blocks: State preparation, a \mbox{problem-specific} oracle, and the diffusor.
To utilize Grover's algorithm for solving a SAT problem, all constraints describing a valid variable assignment must be encoded within the oracle, while the state preparation and diffusor blocks are \mbox{problem-independent}.
Together, the oracle and the diffusor constitute one \emph{Grover iteration}.
Each application of such an iteration increases the probability of obtaining correct solutions from the measurements---a technique called \emph{amplitude amplification}.
If $n$ qubits are used and the underlying problem has $k$ possible solutions, the maximal amplification is achieved after $\lfloor\frac{\pi}{4}\sqrt{\frac{2^n}{k}} \rfloor$ iterations.
\end{example}

\subsection{Encoding}
After an algorithm is selected, the actual problem instance must be encoded in the form of a quantum circuit, such that the chosen quantum algorithm can determine a solution for that specific problem instance.
The encoding fashion highly depends on the chosen algorithm and its inner working mechanisms. 
While some quantum algorithms, such as VQE and QAOA, follow a rather generic general fashion, others, such as Grover's algorithm, are far more \mbox{problem-specific}.

\begin{example}\label{ex:encoding}
In order to apply Grover to the $2\times2$ Kakuro riddle shown in \autoref{fig:wf1}, all three constraints mentioned in \autoref{ex:kakuro} need to be encoded in an oracle as described in \autoref{ex:algo_choice}.
The resulting circuit is shown in \autoref{fig:wf2}.
Since each of the variables only assumes values in $\{0,1\}$, a single qubit can be used to represent each of the variables $a$ to $d$---amounting to four qubits.
The three inequality constraints are then encoded through single CNOT gates (as annotated in \autoref{fig:wf2}).
Additionally, a \emph{flag} qubit is introduced to indicate whether a variable assignment satisfies all constraints.
An MCZ gate (indicated by the four black dots) is used for exactly this purpose and flips the phase of the flag qubit whenever all constraints are satisfied.
Afterwards, all constraints are \emph{uncomputed}---a prerequisite for Grover to function properly. 
With four qubits representing the open variables and two possible solutions, two Grover iterations are necessary (as indicated by the dashed rectangle and \enquote{$\times2$} in \autoref{fig:wf2}).

\end{example}

\subsection{Executing}
Afterwards, the resulting quantum circuit is executed.
In essence, executing a quantum circuit means repeatedly initializing a quantum system (typically to the all-zero state $\ket{0...0}$), applying all the operations of the circuit, and measuring the final state. 
This corresponds to sampling from the probability distribution described by the amplitudes of the final state and can either be performed on a classical computer (using \emph{quantum circuit simulators} such as~\cite{sim1, Zulehner2019AdvancedSO,hillmichJustRealThing2020, vincentJetFastQuantum2021, villalongaFlexibleHighperformanceSimulator2019, qiskit}) or on actual quantum computers (such as from IBM, Rigetti, AQT, Google, Oxford Quantum Computing, or IonQ).
In either case, the result after the execution is a histogram describing the distribution of the measured results.

\begin{example}\label{ex:execution} 
The execution of the circuit from \autoref{ex:encoding} yields an outcome distribution (referred to as a histogram) as illustrated in \autoref{fig:wf3}.
While most of the states have a very low probability of around $0.4\%$,
the two highlighted states ($\ket{0110}$ and $\ket{1001}$) occurred in around $47\%$ of the executions.
\end{example}

\subsection{Decoding}

Based on the histogram obtained from the circuit execution, the solution to the problem must be decoded.
In general, the solutions determined by the quantum algorithm are represented by the bitstrings in the histogram that occurred most frequently.
The amount of post-processing necessary to transform these quantum solutions to classical solutions of the real problem greatly varies depending on the algorithm itself and the complexity of the encoding that was chosen to realize the quantum algorithm.

\begin{example}\label{ex:decoding}
The two bitstrings $0110$ and $1001$ that occurred most frequently in the histogram shown in \autoref{fig:wf3} encode the solutions to the $2\times2$ Kakuro riddle. 
Since every variable has been encoded as a single qubit, it is easy to read out the solutions $a=0$, $b=1$, $c=1$, $d=0$ and $a=1$, $b=0$, $c=0$, $d=1$. 
This eventually lead to the solutions shown in \autoref{fig:wf4}.
\end{example} 

\section{Envisioned Framework}\label{sec:framework}
In this work, we envision a framework for developing quantum computing solutions that aims to \emph{shield} users from the intricacies of quantum computing while at the same time providing them with all that is needed to realize their desired solution. 
To this end, the workflow as reviewed in \autoref{sec:workflow} offers several tangible opportunities:
To begin with, the original problem description is purely classical and, given a suitable quantum algorithm,
its encoding into a quantum circuit can be fully automated.
In a similar fashion, also the execution and decoding step need not be handled by the user but can rather be embedded into an automated workflow as well.
This yields the desired solution which can be returned to the user---again in a classical format.
In doing so, all tedious and \mbox{error-prone} tasks involving huge manual effort are conducted by the framework itself instead of the user.

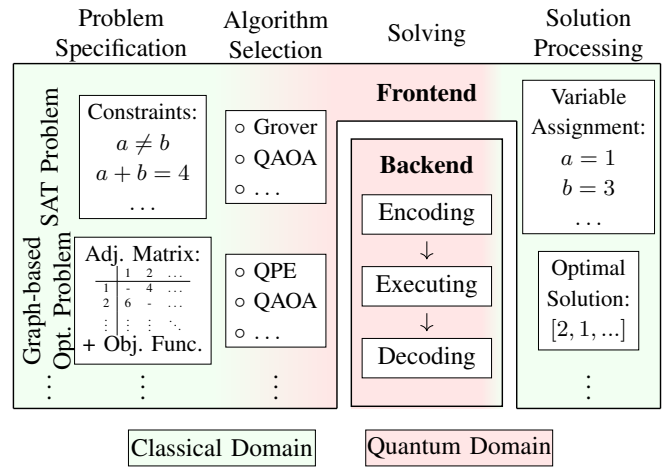
\begin{figure}[t]
	   \resizebox{1.0\linewidth}{!}{
	\begin{tikzpicture}

	\draw [draw=none, fill=green!5!] (-1.5,0-0.4) rectangle (1.3,-5.2);
	\draw [draw=none, left color=green!5!, color=orange, right color=red!10!] (1.3,-0.4) rectangle (3.0,-5.2);
	\draw [draw=none, fill=red!10!] (3.0,-0.4) rectangle (5.0,-1.2);
	\draw [draw=none, left color=red!10!, right color=green!5!] (5.0,-0.4) rectangle (5.3,-1.2);
	\draw [draw=none, fill=green!5!] (5.3,-0.4) rectangle (5.5,-1.2);
	\draw [draw=none, fill=red!10!] (3.2,-1.45) rectangle (5.0,-5.2);
	\draw [draw=none, left color=red!10!, right color=green!5!] (5.0,-1.45) rectangle (5.3,-5.2);
	\draw [draw=none, fill=green!5!] (5.5, -0.4) rectangle (7.5,-5.2);	
	
	\node[align=center] at (0, 0.0) {Problem \\ Specification};		
	\node[align=center, rotate=90] at (-1, -1.7) {SAT Problem};		
	\node[draw,align=center, line width=0.1mm, fill=white] at (0.3, -1.7) {\small{Constraints}:\\  	
	\small{$a\neq b$}\\\small{$a + b =4$}\\ \small{$\hdots$}
	};		
	\node[align=center, rotate=90] at (-1, -3.7) {Graph-based \\ Opt. Problem};
	\node[draw,align=center, fill=white] at (0.3, -3.7) {\small{Adj. Matrix}:\\ 
	\scalebox{0.5}{
		\begin{tabular}{c|ccc}
	 	& 1 & 2 &  $\hdots$ \\ \hline
		1 & - & 4 & $\hdots$ \\
		2 & 6 & - & $\hdots$  \\
		$\vdots$  & $\vdots$  &$\vdots$  &$\ddots$
		\end{tabular} } \\ \small{+ Obj. Func.}
	};		
	
	\node[align=center] at (-1, -4.8) {$\vdots$};		
	\node[align=center] at (0.3, -4.8) {$\vdots$};

	\node[align=center] at (2.15, 0.0) {Algorithm \\ Selection};	
	\node[draw,align=left, fill=white] at (2.15, -1.7) {\small{$\circ$ Grover} \\ \small{$\circ$ QAOA} \\\small{$\circ$ $\hdots$}};		
	\node[draw,align=left, fill=white] at (2.15, -3.7) {\small{$\circ$ QPE} \\ \small{$\circ$ QAOA} \\\small{$\circ$ $\hdots$}};		
	\node[align=center] at (2.15, -4.8) {$\vdots$};		
	
	\node[align=center] at (4.25, 0.0) {Solving};
	\node[align=center] (Backend) at (4.25, -1.8) {{\textbf{Backend}}};	
	\node[align=center] (Backend) at (4.25, -0.8) {{\textbf{Frontend}}};	
	\node[draw,align=center, minimum width = 1.8cm, outer sep=1mm, fill=white] (A) at (4.25, -2.5) {Encoding};	
	\node[draw,align=center, minimum width = 1.8cm, outer sep=1mm, fill=white] (B) at (4.25, -3.5) {Executing};	
	\node[draw,align=center, minimum width = 1.8cm, outer sep=1mm, fill=white] (C) at (4.25, -4.5) {Decoding};	
  	\draw [draw, line width=0.2mm, inner ysep=2mm] (3.2, -1.45) rectangle +(2.1cm, -3.72cm);
	\draw [->] (A) edge (B) (B) edge (C);

	\node[align=center] at (6.5, 0.0) {Solution\\ Processing};
	\node[draw,align=center, fill=white] at (6.5, -1.7) {\small{Variable} \\ \small{Assignment:} \\ \small{$a=1$} \\ \small{$b=3$} \\ \small{$\hdots$}};
	\node[draw,align=center, fill=white] at (6.5, -3.7) {\small{Optimal} \\ \small{Solution:} \\ \small{$ [2,1,...] $ }};		
	\node[align=center] at (6.5, -4.8) {$\vdots$};

	\draw [draw, line width=0.2mm]  (-1.5,-0.4) -- (7.5,-0.4);
	\draw [draw, line width=0.2mm]  (-1.5,-0.4) -- (-1.5,-5.2);
	\draw [draw, line width=0.2mm]  (-1.5,-5.2)-- (3.0,-5.2);
	\draw [draw, line width=0.2mm]  (3.0,-5.2)-- (3.0,-1.2);
	\draw [draw, line width=0.2mm]  (3.0,-1.2)-- (5.5,-1.2);
	\draw [draw, line width=0.2mm] (5.5,-1.2)-- (5.5, -5.2);
	\draw [draw, line width=0.2mm]  (5.5, -5.2)-- (7.5, -5.2);
	\draw [draw, line width=0.2mm]   (7.5, -5.2)-- (7.5, -0.4);

	\draw [align=center, draw, fill=green!5!] (0.1,-5.5) rectangle (2.7,-6.0) node[pos=.5]{Classical Domain};
	\draw [align=center, draw, fill=red!10!] (3.4,-5.5) rectangle (6.0,-6.0) node[pos=.5]{Quantum Domain};
	\end{tikzpicture}
}
	\caption{Envisioned framework.}
	\label{fig:framework}
	\vspace*{-5mm}
\end{figure}

Overall, this leads to a framework as sketched in \autoref{fig:framework} which realizes a quantum computing solution for a given problem in four steps:
\begin{enumerate}
\item \emph{Problem Specification:} The user is asked to classify the problem as well as to insert the particular instance of the problem. To this end, problems are grouped into over-arching problem classes (such as SAT or \mbox{graph-based} optimization). For each problem class, corresponding interfaces are provided that allow the user to specify particular instances of the problem class. 
This is very similar to the inputs of classical solvers and, hence, does not require any quantum expertise at all. 
\item \emph{Algorithm Selection:} The user needs to select which quantum algorithm should be used to solve the problem. While this certainly requires some quantum computing expertise, the framework can actively support the user in this process. In fact, after choosing the problem class, often only a very selected number of algorithms remain appropriate to solve an instance of that class. For example, if a SAT problem has been specified, Grover's algorithm or QAOA seem to be the most promising algorithms. A corresponding \mbox{a-priori} selection can be conducted by quantum computing experts and, afterwards, accordingly incorporated into the framework. By this, the user is properly guided and shielded from the quantum domain as much as possible.
\item \emph{Solving:} With the problem specification and the quantum algorithm set, the quantum computing solution can be realized. This requires the encoding, execute, and decoding steps reviewed above. Interestingly, all those steps can actually be conducted in an automatic fashion and, hence, completely shielded from the user. 
This builds on various demonstrations of problems being translated to the quantum realm, e.g.,~\cite{wf_ex_1, wf_ex_2, wf_ex_3, wf_ex_4}, and defers dealing with the intricacies of quantum computing from the user to the quantum computing expert developing the framework.
\item \emph{Solution Processing:} Finally, the decoded solution is provided to the user in a format as it would have been provided by a classical solver.
\end{enumerate}

This results in a \emph{frontend} relying on classical descriptions and formats, where the user has to provide the problem instance and subsequently receives the solution; and a heavily automated \emph{backend} that takes care of the encoding, executing, and decoding tasks.
This keeps the user (who obviously only interacts with the frontend) in the classical domain (indicated by the green color in \autoref{fig:framework}) and as much as possible shielded from the quantum domain (indicated by the red color in \autoref{fig:framework}).

Eventually, this vision of a framework may enable users to utilize the benefits of quantum computing with no or only moderate background in this technology.
However, implementing such a framework is no easy task: Interfaces, algorithm templates, automatic encoding, and decoding solutions, etc. need to be developed for a broad variety of problem classes.
Moreover, naive implementations will certainly not be sufficient for practical use cases as they would lead to a substantial overhead today's quantum computers cannot handle.
As a consequence, the development of \emph{efficient} encodings and quantum circuit realizations~\cite{zulehnerExploitingCodingTechniques2018, eff2, eff3, eff4, zulehnerOnepassDesignReversible2018, solpaths2022poggel, davio2010, adarshSyReCSynthesizerMQT2022} as well as methods properly compiling the resulting quantum circuits using the most promising compilation options~\cite{quetschlich2022mqtpredictor} have received substantial interest recently.
Overall, providing a fully-fledged implementation of the envisioned framework is clearly out of scope for this work and more of a long-term goal.

However, to allow for initial studies on the principal feasibility of the proposed vision, we realized two \mbox{proof-of-concept} implementations of the envisioned framework: One for the SAT problem class covered above; and one for the class of \mbox{graph-based} optimization problems. %
While these \mbox{proof-of-concept} implementations are certainly not optimized and scalable for practically relevant instances, they showcase that the concepts envisioned above are indeed feasible and can be extended to various problem classes.
However, this comes with a large \mbox{one-time} development effort for each problem class for the quantum computing experts extending the framework.

\section{Proof-of-Concept Implementations}\label{sec:implementation}
In this section, we describe these \mbox{proof-of-concept} implementations. Based on that, \autoref{sec:results} then showcases that a framework as proposed above (even in simple \mbox{proof-of-concept} implementations) indeed allows for a realization of quantum computing solutions with no or only moderate quantum computing expertise.

\subsection{Satisfiability Problems}\label{sec:csp}

As a first proof-of-concept implementation, we considered SAT problems and realized the respective components of the framework that are needed to provide support for this problem class. 
To this end, we created a Python skeleton defining the interfaces sketched in~\autoref{fig:framework}. 
Based on that, the respective components have been realized as follows:

First, we created the respective interface for SAT problems---requiring a dedicated input mask for defining the variables and constraints of arbitrary SAT instances.
This interface is by no means different to the interfaces of classical tools from related domains and communities (e.g., based on SAT~\cite{tseitinComplexityDerivationPropositional1983, biereHandbookSatisfiability2009}, SMT~\cite{demouraZ3EfficientSMT2008, barrettSMTLIBStandardVersion2017, Wille2009SWORD}, or ILP~\cite{nemhauserIntegerCombinatorialOptimization1999}).
Hence, the huge set of description means, parsers, etc. developed in the past can be easily reused.
In this implementation, we eventually decided to provide support for specifying small SAT problems 
formulated as a list of constraints.

Next, templates of possible quantum algorithms suited to tackle the problem are provided. 
As already discussed above in \autoref{ex:algo_choice}, Grover's algorithm is particularly suited for this class of problems while QAOA might also be a possible choice.
For Grover's algorithm, a corresponding template is created, such that the general structure of the quantum algorithm is utilized for the subsequent encoding.
This template consists of a state preparation, an oracle, and a diffuser (building) block.

To fill these blocks, automatic methods have been implemented that take the problem description in the form of the constraints list and construct the corresponding quantum circuit.
More precisely, qubits are allocated for each variable in the constraints.
From that, the state preparation and diffusion building blocks can already be fully instantiated (as they only depend on the number of qubits used).
Then, the main task lies in encoding the given classical constraints in the form of an oracle, i.e., in terms of quantum gates.
Since quantum gates (and quantum computing itself) are inherently reversible---while many classical functions are not---dedicated synthesis techniques need to be applied for that matter~\cite{zulehnerMakeItReversible2017}.
In our \mbox{proof-of-concept} implementation, each constraint is synthesized on a separate qubit employing existing techniques for reversible logic synthesis (such as quantum adders~\cite{rev_adder_1, rev_adder_2, rev_adder_3}).
Then, all constraints are combined using an MCZ gate (similar to the encoding of the Kakuro riddle in \autoref{ex:encoding}).

After constructing the entire circuit from the given problem description in a fully automated fashion, it can now directly be passed to any device capable of its execution---either an actual quantum computer (e.g., from IBM, Rigetti, AQT, Google, Oxford Quantum Computing, or IonQ) or a classical quantum circuit simulator (e.g.,~\cite{sim1, Zulehner2019AdvancedSO, hillmichJustRealThing2020, vincentJetFastQuantum2021, villalongaFlexibleHighperformanceSimulator2019, qiskit}).
For the purpose of this work, we opted for the \emph{MQT DDSIM}-simulator (taken from~\cite{Zulehner2019AdvancedSO}) as a simple and easy to incorporate solution to generate the respective outcome distributions for the constructed circuits.

Since the number of solutions to the original problem might not be known \mbox{a-piori} (and, hence, the optimal number of Grover iterations cannot be statically determined), the proposed implementation dynamically adjusts the number of iterations until a definite result is obtained.
Given the most probable measurement results, the respective variable assignments are then inferred in a similar fashion as in \autoref{ex:decoding}.
Accordingly, the determined variable assignments are returned to the frontend, which visualizes them in a format familiar to the user.

\subsection{Graph-based Optimization Problem: \\ Travelling Salesman Problem}\label{sec:tsp}
To demonstrate the flexibility and expandability of the envisioned framework, a proof of concept for a different problem class (namely \mbox{graph-based} optimization problems) has additionally been implemented.

Analogously to the previous implementation, we designed the respective interface for defining \mbox{graph-based} optimization problems by providing means to specify the \mbox{graph-defining} weighted adjacency matrix and the objective function.
Again, this interface is similar to the interface of classical solvers for this type of problem.

For the proof-of-concept implementation, we particularly considered the \emph{Travelling Salesman Problem} (TSP) as an objective function.
In simple words, solving a TSP means to determine the shortest path visiting all nodes and ending at the starting node while each node is passed exactly once.
More formally, the solution to the TSP is given by the shortest \emph{Hamiltonian cycle} of the graph.
The \emph{Quantum Phase Estimation} (QPE,~\cite{kitaevQuantumMeasurementsAbelian1995}) algorithm has been determined as a suitable technique for solving this problem on a quantum computer based on the encoding technique presented in~\cite{tsp_qpe}.

The QPE algorithm allows to efficiently estimate the phase of one of a unitary matrix $U$'s eigenvalues, i.e., estimates $\theta$ such that $U\ket{\psi}=e^{2\pi i\theta}\ket{\psi}$ for a given eigenstate $\ket{\psi}$.
Similar to Grover's algorithm, where the encoding mainly concerns the design of the oracle part of the algorithm, the essential part of solving a problem with QPE is designing the unitary operator $U$ so that the answer to the problem lies in the phase of its eigenvalues.
For a TSP problem with $N$ nodes, $N\lceil\log_2{N}\rceil$ qubits are allocated to encode all possible Hamiltonian cycles.
Next, the adjacency matrix is encoded by constructing a unitary matrix that encodes the respective weights as phases on its diagonal.
By design, the Hamiltonian cycles of the graph (encoded as states $\ket{\psi}$) are eigenstates of this constructed operator $U$ and the corresponding phase reflects the length of the cycle---the lower the phase, the shorter the cycle.

In the execution step, the QPE algorithm is applied to each unique Hamiltonian cycle---again utilizing the same quantum circuit simulator as before.
Each execution run produces an estimate for the length of the corresponding Hamiltonian cycle.
Accordingly, the path corresponding to the smallest measured phase is returned to the frontend as the solution to the problem---ready to be visualized and shown to the user.

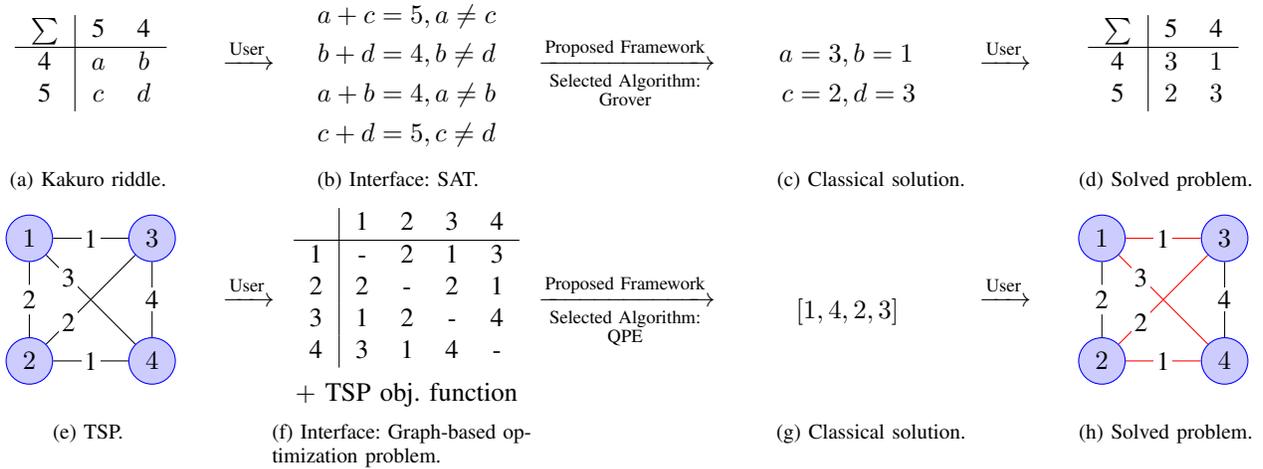
\begin{figure*}[h]
\centering
\begin{subfigure}{.18\linewidth}
    \centering
    	\begin{tabular}{c|cc}
$\sum$  & 5 & 4   \\ \hline
4 & $a$  & $b$   \\ 
5 & $c$  & $d$   \\ 
\end{tabular}
\end{subfigure}
{$\xrightarrow{\text{User}}$}
\begin{subfigure}{.18\linewidth}
    \centering
\begin{align*}
  a + c &= 5, a\neq c  \\
  b + d &= 4, b\neq d  \\
  a + b &= 4, a\neq b  \\
  c + d &= 5, c \neq d 
\end{align*}
\vspace*{-0.5cm}
\end{subfigure}
{$\xrightarrow[\substack{\text{Selected Algorithm:} \\ \text{Grover}}]{\substack{\text{Proposed Framework}}}$}
\begin{subfigure}{.18\linewidth}
    \centering
\begin{align*}
  a &= 3, b = 1 \\
  c &= 2, d = 3
\end{align*}
\vspace*{-0.5cm}
\end{subfigure}
{$\xrightarrow{\text{User}}$}
\begin{subfigure}{.18\linewidth}
    \centering
        	\begin{tabular}{c|cc}
$\sum$  & 5 & 4   \\ \hline
4 & $3$  & $1$   \\ 
5 & $2$  & $3$   \\ 
\end{tabular}
\end{subfigure}

	\subfloat[Kakuro riddle.\label{fig:csp1}]{\hspace{.19\linewidth}}
	\hspace{6mm}
	\subfloat[Interface: SAT. \label{fig:csp2} ]{\hspace{.19\linewidth}}
	\hspace{.15\linewidth}
	\subfloat[Classical solution. \label{fig:csp3} ]{\hspace{.19\linewidth}}
	\hspace{3mm}
	\subfloat[Solved problem.\label{fig:csp4} ]{\hspace{.19\linewidth}}

\begin{subfigure}{.18\linewidth}
    \centering
\begin{tikzpicture}[
   node_style/.style={circle,draw=blue,fill=blue!20!},
   edge_style/.style={draw=black}]
  \node[node_style]         (1)                        {$1$};
  \node[node_style]          (2) [below = of 1]    {$2$};
  \node[node_style]          (3) [right  =of 1]    {$3$};
  \node[node_style]          (4) [right  =of 2]    {$4$};
  
    \draw[edge_style]  (1) edge node[inner sep=1pt, circle, fill=white] {2} (2);
    \draw[edge_style]  (1) edge node[inner sep=1pt, fill=white] {1} (3);
    \draw[edge_style]  (1) edge node[inner sep=1pt, fill=white, pos=0.25]{3} (4);
    \draw[edge_style]  (2) edge node[inner sep=1pt, fill=white, pos=0.25]{2} (3);
    \draw[edge_style]  (2) edge node[inner sep=1pt, fill=white]{1} (4);
    \draw[edge_style]  (3) edge node[inner sep=1pt, fill=white] {4} (4);
  \end{tikzpicture}
\end{subfigure}
{$\xrightarrow{\text{User}}$}
\begin{subfigure}{.18\linewidth}
    \centering
\vspace*{0.2cm}
	\begin{tabular}{c|cccc}
  & 1 & 2 & 3 & 4 \\ \hline
1 & -  & 2   &  1 & 3\\
2 & 2  & -  & 2  & 1  \\
3 & 1  & 2  & -  &  4 \\
4 &  3 & 1  & 4  & -
\end{tabular}

\vspace*{0.2cm}

$ + \text{ TSP obj. function}$
\end{subfigure}
{$\xrightarrow[\substack{\text{Selected Algorithm:} \\ \text{QPE}}]{\substack{\text{Proposed Framework}}}$}
\begin{subfigure}{.18\linewidth}
    \centering
\begin{align*}
 [1,4,2,3]
\end{align*}
\vspace*{-0.5cm}
\end{subfigure}
{$\xrightarrow{\text{User}}$}
\begin{subfigure}{.18\linewidth}
    \centering
\begin{tikzpicture}[
   node_style/.style={circle,draw=blue,fill=blue!20!},
   edge_style/.style={draw=black},
   edge_style_chosen/.style={draw=red}]
  \node[node_style]         (1)                        {$1$};
  \node[node_style]          (2) [below = of 1]    {$2$};
  \node[node_style]          (3) [right  =of 1]    {$3$};
  \node[node_style]          (4) [right  =of 2]    {$4$};
  
    \draw[edge_style]  (1) edge node[inner sep=1pt, circle, fill=white] {2} (2);
    \draw[edge_style_chosen]  (1) edge node[inner sep=1pt, fill=white] {1} (3);
    \draw[edge_style_chosen]  (1) edge node[inner sep=1pt, fill=white, pos=0.25]{3} (4);
    \draw[edge_style_chosen]  (2) edge node[inner sep=1pt, fill=white, pos=0.25]{2} (3);
    \draw[edge_style_chosen]  (2) edge node[inner sep=1pt, fill=white]{1} (4);
    \draw[edge_style]  (3) edge node[inner sep=1pt, fill=white] {4} (4);
  \end{tikzpicture}
\end{subfigure}

	\subfloat[TSP.\label{fig:tsp1}]{\hspace{.19\linewidth}}
	\hspace{6mm}
	\subfloat[Interface: Graph-based optimization problem. \label{fig:tsp2} ]{\hspace{.19\linewidth}}
	\hspace{.15\linewidth}
	\subfloat[Classical solution. \label{fig:tsp3} ]{\hspace{.19\linewidth}}
	\hspace{3mm}
	\subfloat[Solved problem.\label{fig:tsp4} ]{\hspace{.19\linewidth}}

\caption{Case studies: Using the proposed framework to solve two problem instances of different problem classes.}
\label{fig:case_study}
\vspace{-2mm}
\end{figure*}
 
\section{Case Studies: User's Perspective}\label{sec:results}
Eventually, the steps described in the previous section led to a \mbox{proof-of-concept} implementation of the ideas proposed in \autoref{sec:framework} which are publicly available on GitHub (\url{https://github.com/cda-tum/MQTProblemSolver}) as part of the \emph{Munich Quantum Toolkit (MQT)}.
In this section, the resulting framework realization is evaluated from a user's perspective in order to demonstrate that the proposed approach indeed allows one to utilize quantum computing for solving classical problems with no or only moderate quantum computing expertise.
For that, problem instances from both the SAT and the graph-based optimization problem classes are exemplary solved.

\subsection{SAT Problem: Kakuro Riddle}
Consider again the Kakuro riddle that was used as a running example throughout this work.
\autoref{fig:csp1} shows a slightly more complex instantiation of such a riddle.
The goal is to determine values for $a$ to $d$ such that the sums add up to $4$ and respectively $5$.

Using the proposed framework, the user first has to identify the corresponding problem class (here, obviously SAT) and provide the instance.
For the latter, the classical constraints shown in \autoref{fig:csp2} are passed to the framework and Grover's algorithm is selected as the quantum algorithm.
All of that happens on a purely classical basis as it would have if the problem were to be solved using classical solvers.
Afterwards, with the push of a button, the framework determines the variable assignment for $a$ to $d$ shown in \autoref{fig:csp3}---completely shielding the user from all the tedious and error-prone quantum steps.
The solved Kakuro instance is shown in \autoref{fig:csp4}.

\subsection{Graph-based Optimization Problem: TSP}
In a similar fashion, the TSP shown in \autoref{fig:tsp1} can be automatically solved in a push-button fashion using the implemented framework.
Remember, that the goal of the TSP is to determine the shortest path traversing all nodes exactly once and returning to the start.

All that is required from the user is the adjacency matrix of the graph and the selection of TSP as an objective function as shown in \autoref{fig:tsp2}---again sticking to purely classical and common description means.
After selecting the QPE algorithm for solving the problem, with the push of a button, the framework determines the classical solution as reflected in \autoref{fig:tsp3}.
This corresponds to the traversal of the graph as shown in \autoref{fig:tsp4} (indicated by the red edges).

\section{Conclusions}\label{sec:conclusions}
In this work, we envisioned a framework that allows users from various domains with little to no quantum computing expertise to use quantum computing for solving their problems.
To achieve that, we shielded the user from most quantum computing steps and, instead, proposed corresponding automatic solutions for them.
To this end, interfaces are employed that are as similar as possible to the interfaces of classical solvers for the respective problems.
The feasibility and usability of such a framework has been demonstrated by \mbox{proof-of-concept} implementations and corresponding case studies for two different problem classes which are publicly available on GitHub (\url{https://github.com/cda-tum/MQTProblemSolver}).

While these case studies confirmed the potential of the envisioned framework, developing a \mbox{fully-fledged} implementation will require significant further efforts in many dimensions:
Further problem classes need to be supported, generic implementations of further quantum algorithms need to be provided, and highly efficient encoding techniques need to be developed in order to allow the resulting framework to be of practical use.
While this clearly is out of scope for this work and actually constitutes an ambitious \mbox{long-term} goal, this work lays the foundations for the design automation community to work towards this vision---eventually providing a \mbox{low-threshold} solution for utilizing quantum computing in \mbox{real-world} applications without the necessity of being an expert in quantum computing.

\section*{Acknowledgments}
This work received funding from the European Research Council (ERC) under the European Union’s Horizon 2020 research and innovation program (grant agreement No. 101001318), was part of the Munich Quantum Valley, which is supported by the Bavarian state government with funds from the Hightech Agenda Bayern Plus, and has been supported by the BMWK on the basis of a decision by the German Bundestag through project QuaST, as well as by the BMK, BMDW, and the State of Upper Austria in the frame of the COMET program (managed by the FFG).

\printbibliography

\end{document}